\documentclass[journal,onecolumn,comsoc]{IEEEtran}
\usepackage{nopageno}
\IEEEoverridecommandlockouts
\usepackage{times}
\usepackage{amsmath,amssymb}
\usepackage{graphicx}
\usepackage{color}
\usepackage{soul}
\usepackage{enumerate} 
\usepackage[font = footnotesize]{caption}
\usepackage{subcaption}
\usepackage{cite}
\usepackage{multirow,tabularx}
\usepackage{cases}
\usepackage{ifthen}
\usepackage[ruled,vlined,linesnumbered]{algorithm2e}
\usepackage{wasysym}
\usepackage{stackrel}
\usepackage{bm}
\usepackage{algpseudocode}
\usepackage[linesnumbered,ruled,vlined]{algorithm2e}
\SetKwInput{KwInput}{Input}                
\SetKwInput{KwOutput}{Output}
\SetKwInput{KwData}{Initialize}
\SetArgSty{textnormal}

\newcommand{\hide}[1]{\ifthenelse{\boolean{false}}{#1}{}}

\newtheorem{theorem}{{\bf Theorem}}

\newcommand{\qed}{\nobreak \ifvmode \relax \else
      \ifdim\lastskip<1.5em \hskip-\lastskip
      \hskip1.5em plus0em minus0.5em \fi \nobreak
      \vrule height0.75em width0.5em depth0.25em\fi}

\newcommand{\barr}{\begin{array}}
\newcommand{\earr}{\end{array}}

\newcommand{\benum}{\begin{enumerate}}
\newcommand{\eenum}{\end{enumerate}}

\newcommand{\bsp}{\begin{slide*}}
\newcommand{\esp}{\end{slide*}}
\newcommand{\bsl}{\begin{slide}}
\newcommand{\esl}{\end{slide}}

\newcommand{\set}[1]{\{#1\}}

\algrenewcommand\algorithmicrequire{\textbf{Input:}}
\algrenewcommand\algorithmicensure{\textbf{Output:}}

\IEEEoverridecommandlockouts

\usepackage{xcolor,cite,etoolbox}
\makeatletter 
\pretocmd\@bibitem{\color{black}\csname keycolor#1\endcsname}{}{\fail}
\newcommand\citecolor[1]{\@namedef{keycolor#1}{\color{blue}}}
\makeatother

\usepackage{filecontents}

\begin{document}

\title{Accurate Prediction of Nonlinear Distortion of Multi-Carrier Signals}
\author{Cameron~M.~Pike,~\IEEEmembership{Senior~Member,~IEEE},~Brad~Oney,~\IEEEmembership{Member,~IEEE},
~Gabriel~Hepner, and ~Animesh~Yadav,~\IEEEmembership{Senior~Member,~IEEE}
\thanks{C. M. Pike,  B. Oney, and G. Hepner are with GIRD Systems, Inc., Cincinnati, OH 45246 USA (\{cpike,boney,ghepner\}@girdsystems.com). 
        
A. Yadav is with the School of Electrical Engineering and Computer Science, Ohio University, Athens OH 45701 USA (yadava@ohio.edu)}}

\maketitle
\begin{abstract}
Nonlinearities in power amplifiers adversely affect multi-carrier modulation techniques. Accurate prediction of nonlinear distortion is essential for making design trade-offs between output power and network throughput. We use the series form of the characteristic function (ch.f.) method to predict distortion spectra for sparse multi-carrier transmissions. This method results in efficient calculations of individual signal and distortion components. The method is validated both theoretically and practically. Theoretical validation is performed by modeling the signal as a bandpass Gaussian process that is hard limited, and it is shown that the series ch.f. method produces results that are identical with the classical Price's theorem. Practical validation is shown by considering an orthogonal frequency division multiplexing (OFDM) signal with a fragmented spectrum which is then applied to an amplifier driven into compression for which application of Price's theorem is difficult, and the predicted output spectrum corroborates laboratory measurements. Part of the computational efficiency is realized in that the nonlinearity can be expressed as the fast Fourier transform (FFT) of samples of its forward scattering parameter (i.e., S21) or transconductance function (including AM-PM effects), and distortion contributions of the signal can be expressed as numerical autoconvolutions of the clean spectrum. Signal-to-distortion ratio (SDR) can be easily computed and parameterized across variables of interest, such as overdrive level.
\end{abstract}


\section{Introduction}\label{introduction}
Multi-carrier modulation (MCM) techniques are essential building blocks for future wireless networks. The evolution from orthogonal frequency division multiplexing (OFDM) to more versatile techniques, such as orthogonal time-frequency and space (OTFS), reflects the continuous innovation at the physical layer, aiming to unlock new applications and ensure robust wireless connectivity in all scenarios. However, in \textcolor{black}{MCM} techniques, nonlinear distortion from power amplifiers significantly degrades performance, primarily due to the high peak-to-average power ratio (PAPR) character of their waveforms. The impact of nonlinear distortion on bit error rate (BER) performance for the case of OFDM and OTFS is discussed in \cite{ber_ofdm, sharma, ndl_ofdm, error_performance_otfs}. \textcolor{black}{Therefore, characterizing nonlinearity is crucial as it aids in developing countermeasures for various communication systems \cite{10843673, optical_comm_nld}, and device authentication applications through RF fingerprinting \cite{rf_fingerprinting}.} 

In this letter, our focus is on developing and validating an efficient method for accurately predicting the shape and power of the signal and all distortion spectral components at the output of a nonlinear system fed by a multi-carrier signal. These predictions can be used to estimate the impact of clipping on system performance, including signal-to-distortion ratio (SDR) and BER, and thereby inform a trade study to reduce the cost of a power amplifier by relaxing its requirements, or to increase the output power for an existing amplifier, while accounting for the distortion loss in the receiver's link budget. 

Particularly, our \textcolor{black}{statistical signal processing} approach \textcolor{black}{extends} Rice's characteristic function (ch.f.) method \cite[\S4.8]{rice2} by using a Fourier series representation of the nonlinearity \cite{pike,pikeTelegraph}, (contrast \cite{Tahbazalli}). Consequently, distortion terms take the form of autoconvolutions of the signal spectrum, and the total spectrum is found by a weighted sum of signal and distortion terms. While Rice's method  is effective for stochastic processes of any distribution, as the ch.f. of a wide-sense stationary random process always exists \cite[p.72]{peebles}, it requires evaluating many nested improper integrals, which discourages its widespread use. 

Our approach offers several key benefits: i) it is more amenable to numerical evaluation by eliminating the troublesome nested integrals (which also appear in Price and Bussgang\cite{price,brown,voigtlaender,tugfe}), ii) it can be used for multiple signal components, or signals plus channel and/or phase noise, as shown in \cite{pike,pikeTelegraph}, iii) it can be applied to a wide range of systems, including very-high-frequency (VHF), ultra-high-frequency (UHF), and wideband millimeter-wave amplifiers, and iv) it is broadly applicable to various signals and is especially helpful for signals with complicated spectra that are challenging for other methods, whether spectrum-based or discrete inter-modulation product (IMP) accounting \textcolor{black}{and truncated polynomial models of the nonlinearity}, as in \cite{yiming}. The summary of the contributions presented in this letter is as follows:
\begin{itemize}
    \item We theoretically validate the extended ch.f. method introduced in \cite{pike} by showing that it produces results identical to the widely-used Price's theorem \cite{price} for a classical nonlinearity with a Gaussian bandpass signal.
    \item We apply the method in the laboratory to a case of a real OFDM signal and amplifier, and show that the steps involved in the method are straightforward to apply, utilizing instrument measurements of both the amplifier transfer function and the undistorted spectrum of the signal. 
\end{itemize}

The outline of this paper is as follows. Section \ref{problemStatement} presents the signal model and problem statement. Section~\ref{approach} details the extended ch.f. method. Section \ref{hardLimiter} compares the classical result for a hard-limited bandpass Gaussian signal. Section \ref{results} presents and discusses the laboratory-based experimental results. Finally, Section~\ref{conclusion} concludes the paper.

\section{System Model and Problem Statement}
\label{problemStatement}
\subsection{Signal Model}
We model the OFDM signal $s(t)$ as a stationary random process with root-mean-square (RMS) amplitude $\sigma$ and correlation coefficient $\rho(\tau)$, with time lag $\tau$. $\rho(\tau)$ incorporates transmit pulse shape, center frequency, subcarrier occupancy, and relative power levels. For example, if the baseband pulse shape has an autocorrelation function $\phi(\tau)$, and MCM carriers are at center frequencies $f_k$ and amplitudes $A_k$, then $\rho(\tau)\propto\phi(\tau) \sum_k{ A_k\cos{2\pi f_k \tau}}$. As we will show in the subsequent section, there is no need to explicitly specify $\phi(\tau)$ or $f_k$, but we work directly with $\rho(\tau)$ or its transform, the spectrum of the signal. The OFDM signal is then passed through a non-ideal power amplifier or general nonlinear system to obtain the amplified signal $y(t)$ for transmission. Our objective for this system is to accurately compute the spectrum, transmit power, and distortion at the output of the power amplifier. To achieve this, we need to determine the output correlation function, as detailed in the next section.

\vspace{-0.09 in}
\subsection{Problem Statement}
As an analysis aid, we treat the input as two different signals in two different nonlinearities, which we then unify as a special case of a single signal and nonlinearity.  To this end, we begin by expressing two sample functions from the random process $s(t)$ as $s_1(t)$ and $s_2(t)$, 
stationary jointly random processes with zero mean, variances $\sigma_1^2,\, \sigma_2^2$, and correlation coefficient $\rho(\tau)=\mathbb{E}\{s_1(t+\tau)s_2(t)\}/\sigma_1 \sigma_2$. Assuming an optional dc bias $\beta$ on the input signal, the inputs to the nonlinear amplifier are
\begin{equation}
\label{deqn_e2}
x_1(t) =s_1(t)+\beta; \quad x_2(t) =s_2(t)+\beta.
\end{equation}

Next, representing the nonlinear memoryless\footnote{\textcolor{black}{See Section \ref{discussion}}} distortion by $f(x)$ and $g(x)$, the output of the power amplifier is
\begin{equation}
\label{deqn_e1}
y_1(t) =f\left(x_1(t)\right); \quad y_2(t) =g\left(x_2(t)\right).
\end{equation}


The expression for the output correlation can be written by taking the expectation of the product of the outputs as
\begin{eqnarray}
\Psi(\tau)&=&\mathbb{E}\left\{ y_1(t+\tau) \, y_2(t) \right\},\\
&=&\mathbb{E}\left\{ f \left( s_1(t+\tau) +\beta\right) \, g\left( s_2(t) + \beta\right)\right\}.\label{deqn_e3}
\end{eqnarray}

\section{Proposed Methodology}\label{approach}

\textcolor{black}{We introduce the method by stating it as a theorem, the proof of which is provided by derivation in this section. \\
\begin{theorem} 
If the general characteristic function of an input signal ensemble $\set{V(t) \in \mathbb{R} \| \,{|V(t)|<c}}$ is expressed as $\Phi(u,v,\tau)=\mathbb{E}[\exp(juV(t+\tau) + jvV(t))]$, and a linear or nonlinear system is expressed as a Fourier series, 
\begin{equation}
\label{deqn_e4}
p(x) = \lim_{N\to\infty} \sum_{q=-(N-1)/2}^{(N-1)/2} P_q e^{j \frac{q \pi x}{c}}, 
\end{equation}
then the output correlation is  
\begin{equation}
\label{psiTheorem}
\Psi(\tau) =  \lim_{N\to\infty}\sum_{q=-(N-1)/2}^{(N-1)/2} \sum_{r=-(N-1)/2}^{(N-1)/2}\, {P_q P_r} \, \Phi(\frac{q\pi}{c},\frac{r\pi}{c},\tau).
\end{equation}
\end{theorem}
}

\begin{IEEEproof}
In practical situations, the input signal amplitude is finite, so it is entirely valid to define the nonlinearity in a more convenient form as long as it corresponds (i.e., is congruent) with the actual nonlinearity 
over the range of the input signal \textcolor{black}{whose magnitude is less than $c$}. Following this line of thought, we assume that $f,g$ are periodic functions of the input variable (i.e. $x$, voltage or current) with period $2c$, and can be expressed as Fourier series \eqref{deqn_e4} where, $P_q = \frac{1}{2c} \int_{-c}^{c} p(x) e^{-j\frac{q \pi x}{c}} dx$. \textcolor{black}{Convergence of the Fourier series representation of a function $p(x)$ is \emph{absolute} and \emph{uniform} with respect to $x$ on that interval \cite[p. 106]{churchill} under the conditions that $p$ is continuous, and $p'$ is piecewise continuous, which are met by a physically realizable device. The limit to infinity is implied in \eqref{deqn_e5} and following.} 
Substituting (\ref{deqn_e4}) into (\ref{deqn_e3}) for $f$ and $g$, we get 
\begin{equation}
\label{deqn_e5}
\Psi(\tau) = \mathbb{E} \left\{ \left( \displaystyle\sum_{q=-(N-1)/2}^{(N-1)/2} F_q e^{j \frac{q\pi}{c} [s_1(t+\tau)+\beta]}\right) \right. \times \left. \left( \displaystyle\sum_{r=-(N-1)/2}^{(N-1)/2} G_r e^{j \frac{r\pi}{c} [s_2(t)+\beta]}\right) \right\}.
\end{equation}
After rearranging \eqref{deqn_e5} by removing non-random factors from the expectation we get
\begin{equation}
\label{deqn_e6}
\Psi(\tau) =  \displaystyle\sum_{q=-(N-1)/2}^{(N-1)/2} \sum_{r=-(N-1)/2}^{(N-1)/2}\, \overbrace{F_q G_r}^\text{nonlinearity} \,\overbrace{e^{j(\frac{q\pi}{c} + \frac{r\pi}{c})\beta}}^\text{ch.f. $\Phi_\beta(\textcolor{black}{\frac{q\pi}{c},\frac{r\pi}{c}})$ of bias}
\times \underbrace{\mathbb{E} _{s_1,s_2}\left\{ e^{j \frac{q\pi}{c} s_1(t+\tau)} e^{j \frac{r\pi}{c} s_2(t)} \right\}.}_\text{ch.f. $\Phi_{s_1 s_2}(\textcolor{black}{\frac{q\pi}{c},\frac{r\pi}{c}, \tau})$ of signal}
\end{equation}
The expectation operation on the part of (\ref{deqn_e6}) containing the signal is immediately recognized as the joint characteristic function of $s_1$ and $s_2$. 
(We will henceforth omit $\tau$ from the argument lists for brevity sake.) \eqref{deqn_e6} is valid for signals $s_1(t)$ and $s_2(t)$ of any distribution, including Gaussian. \textcolor{black}{Since the general ch.f. of a sum of r.v.s is the product of the individual ch.f.s \cite[\S 4.8]{rice2}, we recognize that \eqref{deqn_e6} is a restatement of \eqref{psiTheorem}.}
\end{IEEEproof}
If we model the OFDM as Gaussian \textcolor{black}{\cite{araujo}} (i.e., jointly Gaussian random variables), we can write the ch.f. as \cite[page 336]{peebles} 
\begin{equation}
\label{deqn_e7}
\Phi_{s_1 s_2}\Big(\textcolor{black}{\frac{q\pi}{c},\frac{r\pi}{c}}\Big)= e^{-\frac{\pi^2}{2c^2} [\sigma_1^2 q^2 + 2\sigma_1 \sigma_2 qr\rho + \sigma_2^2 r^2]}
= e^{-\frac{\pi^2}{2c^2} (\sigma_1^2q^2+\sigma_2^2r^2)} \sum_{k=0}^{\infty} \frac{(-1)^k \left(\frac{\pi}{c}\right)^{2k} (\sigma_1 \sigma_2 qr\rho)^k}{k!}.
\end{equation}

\textcolor{black}{Non-Gaussian} random variable distributions will have a different appearance of (\ref{deqn_e7}), such as random-phase sinusoid \cite{pike} or random telegraph noise \cite{pikeTelegraph}. We now substitute (\ref{deqn_e7}) into (\ref{deqn_e6}), and collect terms into separate sums over $q$ and $r$.
\begin{equation}
\label{deqn_e8}
\Psi = \sum_{k=0}^{\infty} 
\left[  \sum_{q=-(N-1)/2}^{(N-1)/2} F_q  e^{j\frac{q\pi\beta}{c}} e^{-\frac{\pi^2}{2c^2} \sigma_1^2 q^2} \left( \frac{q\pi}{c} \right)^k \right] \times  \left[  \sum_{r=-(N-1)/2}^{(N-1)/2} G_r e^{j\frac{r\pi\beta}{c}} e^{-\frac{\pi^2}{2c^2} \sigma_2^2 r^2} \left( \frac{r\pi}{c} \right)^k \right] 
\frac{(-1)^k}{k!} (\sigma_1 \sigma_2 \rho)^k
\end{equation}
We observe that this expression decomposes the correlation function $\Psi$ into a power series of the correlation function $\rho(\tau)$ of the original OFDM. 
We now narrow this result to the special case $F_\lambda=G_\lambda$, $\sigma_1=\sigma_2=\sigma$, and $\rho(\tau) $ as the autocorrelation coefficient of the input OFDM process. We then rewrite (\ref{deqn_e8}) as a polynomial in $\rho$ with constants $h_k$ as
\begin{equation}
\label{deqn_e9}
\Psi= \sum_{k=0}^{\infty} h_{k}^2 \frac{(\sigma^2 \rho)^k}{k!},
\end{equation}
\begin{equation}
\label{deqn_e10}
h_{k} =
(j)^{k}  \sum_{\lambda=-(N-1)/2}^{(N-1)/2} F_\lambda \underbrace{e^{j\frac{\lambda\pi\beta}{c}}}_\text{bias}  
\,\underbrace{e^{-\frac{\pi^2}{2c^2}\sigma^2 \lambda^2} \left( \frac{\lambda \pi}{c} \right)^k}_\text{signal},
k \ge 0.
\end{equation}
\textcolor{black}{Although \eqref{deqn_e4} converges uniformly, when $F_\lambda$ are combined with other terms in \eqref{deqn_e10}, the magnitude of $h_k$ may oscillate as it converges; for this reason, formal bounds on $N$ are difficult to derive, and is an open research topic. However, $N$ can be determined empirically such that the incremental contribution of the $N$th term is smaller than the sum of the first $N-1$ terms by a desired amount; e.g. five orders of magnitude will produce -100 dB error in the calculation of \eqref{deqn_e10}}.

\begin{figure}[h]
\centering
\includegraphics[width=3.5in]{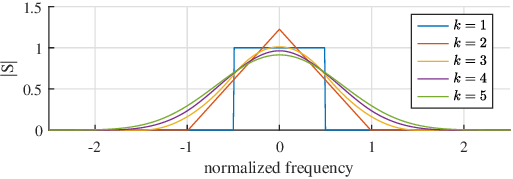}
\caption{Spectra, \textcolor{black}{S,} of the $\rho^k$ contributions in (\ref{deqn_e9}) for $k=1,2,...5$, based on the ideal bandlimited input spectrum of a normalized bandwidth.}
\label{fig1}
\end{figure}

\textit{Remark 1}: Since the spectrum is the Fourier transform of the correlation, the convolution theorem of Fourier transforms tells us that the spectrum of $\rho^{k}$ is the $k$th autoconvolution of the input spectrum, as illustrated in Fig. \ref{fig1}, for an ideal bandlimited process. \eqref{deqn_e9} produces an output spectrum consisting of the sum over $k$ of the signal spectrum convolved with itself $k-1$ times weighted by $h_{k}^2 \sigma^{2k} /k!$. 
This forms the theoretical basis for the specific applications described in the next Sections.

\textit{Remark 2}: It is also observed that \eqref{deqn_e10} facilitates computation of overall signal-to-distortion ratio (SDR) by taking the ratio (linear) or difference (logarithmic) of the power in the signal term for $\rho^1$ to that of the distortion, which is the sum of all the other $\rho^k$. This metric is computed from the nonlinearity alone.
\begin{equation}
\label{deqn_e10a}
\text{SDR} = 10\log_{10}h_1^2\sigma^2 - 10\log_{10}\sum_{k=3,5,...}^\infty \frac{h_k^2 \sigma^{2k}}{k!}
\end{equation}
\vspace{-0.1 in}

\section{Hard Limiter}
\label{hardLimiter}
\begin{figure}[h]
\centering
\includegraphics[width=3.5in]{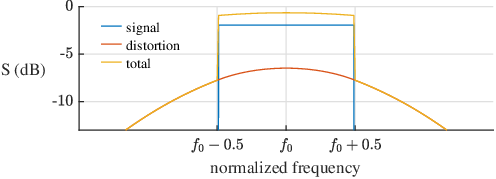}
\caption{Output spectrum, \textcolor{black}{S,} and separated signal and distortion spectra based on the ideal bandlimited input spectrum of a normalized bandwidth, computed according to (\ref{deqn_e9}).}
\label{fig2}
\end{figure}

In this section, we validate the Fourier series approach with a known result from the literature which the reader can easily reproduce, namely, when the nonlinearity is a hard limiter, $f(x)=\text{sgn}(x)$, and the random variable is Gaussian. This scenario is addressed by Price's theorem, where the known output is given by \cite{price}\cite[\S 4.4.40]{abramowitz}
\begin{equation}
\label{deqn_e11}
\Psi=\frac{2}{\pi}\arcsin \rho(\tau) = \frac{2}{\pi} \left( \rho + \frac{\rho^3}{6}+\frac{3\rho^5}{40}+\frac{5\rho^7}{112}\cdots \right).
\end{equation}
When comparing (\ref{deqn_e11}) with (\ref{deqn_e9}) term by term, we observe that constants $h_k^2\sigma^{2k}/k!$ must equal the polynomial coefficients in \eqref{deqn_e11} (including for even $k$, which must all be zero). To verify this, we construct a convenient periodic extension of the signum function as a  ``square function" of the input voltage or current. We select $c=1$, $\beta=0$, and $\sigma=0.1 \ll c$ in (\ref{deqn_e9}) and (\ref{deqn_e10}). It is easy to show that the Fourier series coefficients are
\begin{equation}
\label{deqn_e12}
F_\lambda = 
\begin{cases}
0, &\lambda \text{ even} \\
-\frac{j2}{\lambda \pi}, &\lambda \text{ odd.}
\end{cases}
\end{equation}
Because $f(x)$ is an odd function and the signal is zero mean, only odd powers of $\rho(\tau)$ contribute. (This is also true in the laboratory experiment discussed in the next section.) 
When (\ref{deqn_e12}) is substituted into (\ref{deqn_e10}) and numerically evaluated, a value of $N$ can be found that produces agreement with (\ref{deqn_e11}) within machine precision (including even powers which must be zero). When we apply (\ref{deqn_e9}) and (\ref{deqn_e10}) as weights to an ideal bandlimited signal in Fig. \ref{fig1}, we compute the output spectrum shown in Fig. \ref{fig2}, which agrees with the shape of the spectrum derived in \cite[Fig. 3]{cahn}.  Thus, the series ch.f. method is validated theoretically. The next section shows practical validation.

\section{Results and Discussion}\label{results}
\subsection{Laboratory Experiments}
\label{ofdm}
\begin{figure}[h] 
\centering
\includegraphics[width=3.5in]{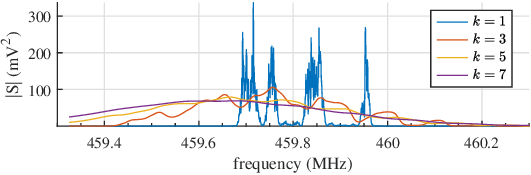}
\caption{Spectra of the $\rho^k$ contributions in (\ref{deqn_e9}) for $k=1,3,5,7$, based on the fragmented input spectrum measured in the laboratory, shown \textcolor{black}{as} $k=1$. \textcolor{black}{Converted to mV\textsuperscript{2} from dBm into 50 ohms. Resolution and video bandwidth were set to 1 kHz, and sample spacing in the data set was an average of 305.17 Hz.}}
\label{fig5}
\end{figure}

In this section, we validate the effectiveness of the method described above in the laboratory using measurements collected from a GIRD OFDM transmitter\footnote{https://www.girdsystems.com/products} and Mini-Circuits ZFL-1000VH\footnote{https://www.minicircuits.com/WebStore/dashboard.html?model=ZFL-1000VH} amplifier at 460 MHz. 
In the laboratory, we can directly measure an amplifier's nonlinearity with a vector network analyzer (VNA), in the form of forward scattering parameter S\textsubscript{21} as a function of input signal power. We can also measure a spectrum directly with a spectrum analyzer. Measurements are available from both instruments as comma-delimited text files. To showcase the accuracy of the proposed method with a complicated spectrum, we create an OFDM signal with 16 subcarriers, of which half are turned off\footnote{\textcolor{black}{This is a multiuser system wherein subsets of users are assigned to several subcarriers, and unused subcarriers are turned off.}}, thereby producing a fragmented spectrum (Fig. \ref{fig5}, $k=1$). The active subcarriers are stimulated with random QPSK data streams, and the combination is effectively Gaussian by the central limit theorem \textcolor{black}{and the findings of \cite{araujo}}. 

The steps to compute the output spectrum are as follows: 

\begin{enumerate}[(a)]
    \item begin by deriving the Fourier series coefficients from VNA measurements of the amplifier gain curve,
    \item next, calculate the odd-order autoconvolutions of the input spectrum (from the spectrum analyzer), 
    \item finally, combine these results into the output spectrum using equations (\ref{deqn_e9}) and (\ref{deqn_e10}).
\end{enumerate}

\begin{figure}[!t]
\centering
\includegraphics[width=3.5in]{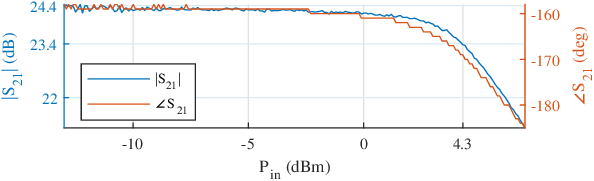}
\caption{Measured S\textsubscript{21} for the ZFL-1000VH at 460 MHz, \textcolor{black}{swept in power from -13 dBm to +7 dBm (0.07 V\textsubscript{p} to 0.7 V\textsubscript{p})}. The 1-dB compression point is marked at 4.3 dBm input power.}
\label{fig3}
\end{figure}

Fig. \ref{fig3} plots the magnitude $|\text{S}\textsubscript{21}|$ and phase $\angle{\text{S}\textsubscript{21}}$ of the gain of the ZFL-1000VH measured in the laboratory. The non-constant phase of S\textsubscript{21} is due to the AM-PM conversion phenomenon (see Section \ref{discussion}). In the figure, the one-dB compression point 
(P1dB) is indicated at +4.3 dBm.

\begin{figure}[!t]
\centering
\includegraphics[width=3.5in]{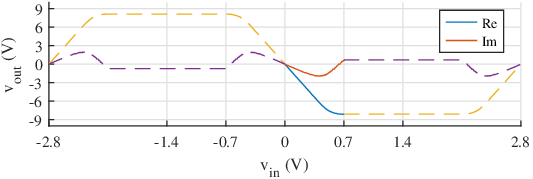}
\caption{Voltage transfer function (solid traces, derived from Fig. \ref{fig3}) and periodic extension (dashed) for the ZFL-1000VH at 460 MHz.}
\label{fig4}
\end{figure}

The first step is to obtain a Fourier series representation of the nonlinearity from the data of Fig. \ref{fig3}. This is accomplished by deriving a voltage transfer characteristic by first computing output power P\textsubscript{out} versus P\textsubscript{in}, then converting the output dBm and phase to the complex voltage relationship shown with the solid curves in Fig. \ref{fig4}. A smooth periodic extension is then devised as shown in dashed traces with period $2c=5.6\text{V}$, \textcolor{black}{a convenient multiple of the peak input voltage in Fig. \ref{fig3},} that has an abundance of ``headroom" above the compression point to allow for large excursions of the input signal. 

We compute the Fourier series coefficients from the samples of v\textsubscript{out} over the interval of v\textsubscript{in} from $0$ to $5.6\text{V}$, using the actual samples from the solid curve appropriately shifted and reflected. For $N$ points FFT and v\textsubscript{out} measurements spaced at $\text{v}_\text{in}(n)=5.6n/N$ volts, we approximate the Fourier series coefficients as $\{F_\lambda\} \approx \frac{1}{N}\text{FFT}_N \{\text{v}_\text{out} (n) \}.$

We proceed to derive odd orders of autoconvolutions of the input spectrum. Rather than building an analytical model of the signal or its spectrum, we can go directly to a measured ``clean" spectrum of the input signal. This is a sample function of the r.v. in the form of its spectrum.
By successive uses of MATLAB's {\tt{conv(...,`same')}} function, we compute the odd-order autoconvolutions of the spectrum, several of which are also plotted in Fig. \ref{fig5}. These are the transforms of the $\rho^k$ in (\ref{deqn_e9}).

\begin{figure}[!t]
\centering
\includegraphics[width=3.5in]{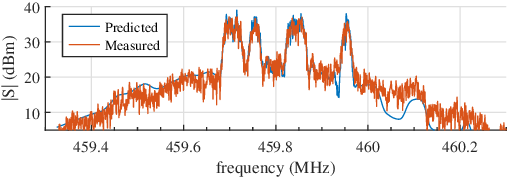}
\caption{Predicted and measured output spectrum for input power level of +4.5 dBm, showing close agreement. \textcolor{black}{Using $N=8001$.}}
\label{fig6}
\end{figure}

 With all the necessary information at hand, including the Fourier series coefficients and the autoconvolutions of the input spectrum, we can predict the output spectrum for a +4.5 dBm input signal. The RMS voltage of this input signal is 0.38V, indicating that the peak voltage is well within the compression region shown in Fig. \ref{fig4}, close to \(3\sigma \approx 1.13\text{V}\). We compute the weighted sum of the autoconvolutions of the signal spectrum, according to the weights calculated in (\ref{deqn_e10}). The resulting spectrum is plotted in Fig. \ref{fig6}, along with actual measurements from the spectrum analyzer for this condition, showing very close agreement, even to the shape of in-band and out-of-band distortion.

\begin{figure}[!t]
\centering
\includegraphics[width=3.5in]{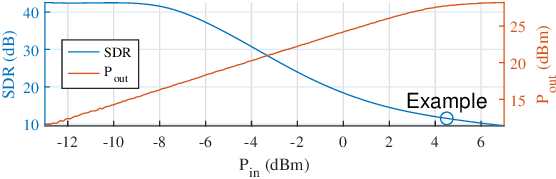}
\caption{Predicted SDR for this amplifier with Gaussian distributed input.}
\label{fig7}
\end{figure}

Fig.~\ref{fig7} plots SDR alongside the output power for different values of input power. The information plotted in this figure can be used in a tradeoff among overdrive level, output power, and SDR. The overall SDR is calculated directly from the coefficients by (\ref{deqn_e10a}) which for this example is 11.5 dB. As seen in (\ref{deqn_e10}) and (\ref{deqn_e10a}), the SDR depends only upon the nonlinear function $f(x)$ and signal power $\sigma^2$ (and indirectly, on the Gaussian distribution of the input signal), so can be predicted for the entire range of input power of interest without having to set up this measurement in the laboratory.  

\subsection{Observations and Discussion }
\label{discussion}
\noindent  The laboratory-measured output spectrum consists of an inseparable mixture of signal and distortion. However, the method presented allows for their clear separation, as shown in Fig. \ref{fig5}. This separation enables an engineering tradeoff of PAPR vs. packet error rate by predicting specific distortion powers in subchannels as the amount of input overdrive is changed. Of note is that the last example did not rely on models of the nonlinearity or the signal correlation function, but we used direct laboratory measurements for both and processed them according to the theoretical framework. The only model was that of the statistical distribution of the signal, namely Gaussian, even though the actual message data was scrambled QPSK on each active OFDM subchannel. The Gaussian model factored into the weights (\ref{deqn_e10}) that dictate the contributions of the several distortion terms. 
While the foregoing derivation of (\ref{deqn_e10}) may be used verbatim for many OFDM applications, a signal of a different distribution, such as PAPR-reduced precoded OTFS \cite{gao}, would necessitate a different appearance of (\ref{deqn_e10}), as shown in \cite{pike,pikeTelegraph}. Derivation of suitable generalized ch.f.s identified in (\ref{deqn_e6}) for other distributions of interest is an ongoing topic of research.

\textcolor{black}{The foregoing analysis and examples assumed a memoryless nonlinearity. All practical devices exhibit memory, often manifesting as amplitude-dependent phase shift and frequency-dependent phase shift. The former (AM to PM conversion) automatically factored into the presented method, as seen in the non-constant phase shift of Fig. \ref{fig3}. The latter phenomenon was not a factor in the lab experiment due to the narrow bandwidth of the signal. Extending this method to wider band systems is an open research topic.}

\section{Conclusion}\label{conclusion}
In this work, we validated the extended ch.f. method, both theoretically by comparison with a classical hard limiter, and experimentally in the laboratory. The ease of calculations for the laboratory example is evident due to the use of the FFT and the compact expressions of (\ref{deqn_e9}) and \eqref{deqn_e10}. Adaptation to real-time performance or use in a predistortion mechanism is a topic of future research, \textcolor{black}{as is application to other systems, such as cartesian nonlinearities}. From a wider perspective, the foregoing procedure is readily applicable to multiple signals of differing distributions (whether Gaussian or not) and spectra, and having been validated, the method can be used with greater confidence in its accuracy to predict distortion spectrum and power. Moreover, this method simplifies the computation of SDR, a crucial metric for design engineers when selecting amplifiers. \textcolor{black}{MATLAB code is provided as a supplement.}

\bibliographystyle{IEEEtran}
\bibliography{myreferences.bib}

\begin{thebibliography}{10}
\providecommand{\url}[1]{#1}
\csname url@samestyle\endcsname
\providecommand{\newblock}{\relax}
\providecommand{\bibinfo}[2]{#2}
\providecommand{\BIBentrySTDinterwordspacing}{\spaceskip=0pt\relax}
\providecommand{\BIBentryALTinterwordstretchfactor}{4}
\providecommand{\BIBentryALTinterwordspacing}{\spaceskip=\fontdimen2\font plus
\BIBentryALTinterwordstretchfactor\fontdimen3\font minus \fontdimen4\font\relax}
\providecommand{\BIBforeignlanguage}[2]{{%
\expandafter\ifx\csname l@#1\endcsname\relax
\typeout{** WARNING: IEEEtran.bst: No hyphenation pattern has been}%
\typeout{** loaded for the language `#1'. Using the pattern for}%
\typeout{** the default language instead.}%
\else
\language=\csname l@#1\endcsname
\fi
#2}}
\providecommand{\BIBdecl}{\relax}
\BIBdecl

\bibitem{ber_ofdm}
T.~Helaly, R.~Dansereau, and El-Tanany, ``{BER} performance of {OFDM} signals in presence of nonlinear distortion due to {SSPA},'' \emph{Wireless Pers. Commun.}, vol.~64, no.~6, pp. 749--760, Jun. 2012.

\bibitem{sharma}
S.~Sharma, A.~Singh, K.~Deka, and C.~Adjih, ``Impact of nonlinear power amplifier on {BER} performance of {OTFS} modulation,'' in \emph{Proc. IEEE Int. Conf. Adv. Netw. Telecommun. Syst. (ANTS)}, Jaipur, India, 17-20 Dec. 2023, pp. 627--632.

\bibitem{ndl_ofdm}
Y.~Du, L.~Hao, Y.~Lei, Q.~Yang, and S.~Xu, ``Nonlinear multicarrier transmitter system with signal clipping: Measurement, analysis, and optimization,'' \emph{IEEE Systems J.}, vol.~18, no.~2, pp. 1426--1435, Jun. 2024.

\bibitem{error_performance_otfs}
S.~G. Neelam and P.~R. Sahu, ``Error performance of {OTFS} in the presence of {IQI} and {PA} nonlinearity,'' in \emph{Proc. 2020 Nat. Conf. Commun (NCC)}, IIT Kharagpur, India, 21-23 Feb. 2020, pp. 1--6.

\bibitem{10843673}
Z.~Mokhtari, R.~Dinis, S.~Hu, and D.~Kapetanovic, ``Joint channel and nonlinearity estimation for memoryless nonlinear systems,'' \emph{IEEE Access}, vol.~13, pp. 13\,143--13\,155, 2025.

\bibitem{optical_comm_nld}
J.~Li, T.~Ye, X.~Su, Y.~Fan, K.~Zhang, S.~Shi, H.~Nakashima, T.~Hoshida, Z.~Tao, Y.~Mori, and H.~Hasegawa, ``Deep-depth probability-maintained notch enabling nonlinear distortion measurement for various modulation formats,'' \emph{J. Lightwave Technol.}, vol.~43, no.~10, pp. 4722--4730, May 2025.

\bibitem{rf_fingerprinting}
W.~Jing, L.~Peng, J.~Zhang, and H.~Fu, ``An investigation of power amplifier feature for deep learning based {RF} fingerprint identification,'' in \emph{Proc. IEEE INFOCOM WKSHPS}, Vancouver, Canada, May 2025, pp. 1--6.

\bibitem{rice2}
S.~O. Rice, ``Mathematical analysis of random noise,'' \emph{Bell Syst. Tech. J.}, vol.~24, no.~1, pp. 46--156, Jan. 1945.

\bibitem{pike}
C.~Pike, ``Fourier series characteristic function method of analysis of sinusoids plus noise in nonlinear devices,'' \emph{TechRxiv.}, January 26 2024.

\bibitem{pikeTelegraph}
C.~M. Pike, ``Analysis of sinusoid plus random telegraph noise in nonlinear devices: Series characteristic function method,'' \emph{IEEE Trans. Circuits Syst. II: Express Br.}, vol.~71, no.~10, pp. 4536--4540, Oct. 2024.

\bibitem{Tahbazalli}
P.~Tahbazalli and A.~Liscidini, ``Gain and nonlinearity analysis in quantized-analog amplifiers using fourier transform,'' \emph{IEEE Transactions on Circuits and Systems I: Regular Papers}, pp. 1--12, 2025.

\bibitem{peebles}
J.~P.~Z. Peebles, \emph{Probability, random variables, and random signal principles}.\hskip 1em plus 0.5em minus 0.4em\relax McGraw-Hill, 2nd Edition, 1987.

\bibitem{price}
R.~Price, ``A useful theorem for nonlinear devices having {G}aussian inputs,'' \emph{IRE Trans. Inf. Theory}, vol.~4, no.~2, pp. 69--72, Jun. 1958.

\bibitem{brown}
J.~Brown, ``Generalized form of {P}rice's theorem and its converse,'' \emph{IEEE Trans. Inf. Theory}, vol.~13, no.~1, pp. 27--30, Jan. 1967.

\bibitem{voigtlaender}
F.~Voigtlaender, ``A general version of price’s theorem,'' \emph{J Theor Probab}, vol.~34, p. 1474–1485, 2021.

\bibitem{tugfe}
O.~T. Demir and E.~Bjornson, ``The {B}ussgang decomposition of nonlinear systems: Basic theory and {MIMO} extensions [{L}ecture {N}otes],'' \emph{IEEE Signal Process. Mag.}, vol.~38, no.~1, pp. 131--136, Jan. 2021.

\bibitem{yiming}
L.~Yiming and M.~O'Droma, ``A novel decomposition analysis of nonlinear distortion in {OFDM} transmitter systems,'' \emph{IEEE Trans. Signal Process.}, vol.~63, no.~19, pp. 5264--5273, 2015.

\bibitem{churchill}
R.~V. Churchill and J.~W. Brown, \emph{Fourier series and boundary value problems}.\hskip 1em plus 0.5em minus 0.4em\relax Third edition, McGraw-Hill, 1978.

\bibitem{araujo}
T.~Araujo and R.~Dinis, ``On the accuracy of the {G}aussian approximation for the evaluation of nonlinear effects in {OFDM} signals,'' \emph{IEEE Trans. Commun.}, vol.~60, no.~2, pp. 346--351, Feb. 2012.

\bibitem{abramowitz}
M.~Abramowitz and I.~Stegun, \emph{Handbook of Mathematical Functions with Formulas, Graphs, and Mathematical Tables}.\hskip 1em plus 0.5em minus 0.4em\relax Dover Publications, Inc.31 E. Second St. Mineola, NYUnited States, 1974.

\bibitem{cahn}
C.~Cahn, ``A note on signal-to-noise ratio in band-pass limiters,'' \emph{IRE Trans. Inf. Theory}, vol.~7, no.~1, pp. 39--43, Jan. 1961.

\bibitem{gao}
S.~Gao and J.~Zheng, ``Peak-to-average power ratio reduction in pilot-embedded {OTFS} modulation through iterative clipping and filtering,'' \emph{IEEE Commun. Lett.}, vol.~24, no.~9, pp. 2055--2059, Sep. 2020.

\end{thebibliography}
\end{document}